\documentclass{article}

\usepackage{fullpage}
\usepackage{inputenc}
\usepackage{fullpage}
\usepackage[colorlinks]{hyperref}
\hypersetup{
     colorlinks   = true,
     linkcolor    = blue,
     urlcolor     = blue,
     citecolor    = blue
}
\usepackage{graphicx}
\usepackage{subcaption} 
\usepackage{listings}
\usepackage{url}
\usepackage{comment}
\usepackage{fancybox}
\usepackage{xcolor}
\usepackage{algorithm}
\usepackage{algpseudocode}
\usepackage{enumitem}  
\usepackage{amsthm}

\lstdefinestyle{code_style}{
    basicstyle=\ttfamily\footnotesize,
    breakatwhitespace=false,         
    breaklines=true,                 
    captionpos=b,                    
    keepspaces=true,                 
    numbers=left,                    
    numbersep=5pt,                  
    showspaces=false,                
    showstringspaces=false,
    showtabs=false,                  
    tabsize=2
}

\title{Collective Vector Clocks: Low-Overhead Transparent Checkpointing for MPI}

\author{
Yao Xu \\
Khoury College of Computer Sciences \\
Northeastern University\\
Boston, MA, USA \\
xu.yao1@northeastern.edu
\and
Gene Cooperman \\
Khoury College of Computer Sciences \\
Northeastern University\\
Boston, MA, USA \\
gene@ccs.neu.edu
}

\date{}

\begin{document}

\maketitle

\begin{abstract}
Taking snapshots of the state of a distributed computation is useful
for off-line analysis of the computational state,
for later restarting from the saved snapshot,
for cloning a copy of the computation, and
for migration to a new cluster.
The problem is made more difficult when supporting collective operations
across processes, such as barrier, reduce operations, scatter and gather, etc.
Some processes may have reached the barrier or other collective operation,
while other processes wait a long time to reach that same barrier or collective operation.
At least two solutions are well-known in the literature:
(i)~draining in-flight network messages and then freezing the network
at checkpoint time; and
(ii)~adding a barrier prior to the collective operation, and either
completing the operation or aborting the barrier if not all
processes are present.
Both solutions suffer important drawbacks.
The code in the first solution must be updated whenever one ports to
a newer network.
The second solution implies additional barrier-related network traffic
prior to each collective operation.
This work presents a third solution that avoids both drawbacks.
There is no additional barrier-related traffic, and the solution
is implemented entirely above the network layer.
The work is demonstrated in the context of transparent checkpointing
of MPI libraries for parallel computation, where each of the first two
solutions have already been used in prior systems, and then abandoned
due to the aforementioned drawbacks.  Experiments demonstrate the
low runtime overhead of this new, network-agnostic approach.
The approach is also extended to non-blocking, collective operations
in order to handle overlapping of computation and communication.
\end{abstract}

\section{Introduction}
\label{sec:introduction}


Collective operations are critically important both for distributed computing
and for parallel computing.  A call to a collective operation involves a fixed set of participating
processes.
Typical collective operations across multiple
computer nodes include: Barrier;
Bcast (Broadcast); Reduce (by a commutative, associative operation such
as sum or max); and Alltoall.  Such collective operations can come in two
``flavors'':  blocking and non-blocking.  Blocking operations allow communication by
multiple processes in order to cooperate in a larger task, and non-blocking operations
allow those same processes to overlap computation and communication.

Calls to the non-blocking versions are completed in two phases.
First, each participating
process must (a)~initiate the collective call; and (b)~later \emph{complete} the
collective operation.  The completion of a collective call can be implemented in one of two ways.
Either the initial call create a \emph{request object} that can later
be polled to test for completion, or else the initial call registers a callback
function that will be called asynchronously by the system when the
operation is complete.

There is a tension between support for collective operations (both blocking and non-blocking)
and a common requirement for transparent checkpointing of
a distributed or parallel computation.  This work concentrates on the domain
of parallel computing, and in particular, MPI (Message Passing Interface).
For each distributed operation described here as \texttt{xxx}, an MPI implementation
defines \texttt{MPI\_Xxx} (for example, \texttt{MPI\_Barrier}, \texttt{MPI\_Bcast}, etc.).  Similarly,
a nonblocking variant is defined by \texttt{MPI\_Ixxx} (for example, \texttt{MPI\_Ibarrier}, \texttt{MPI\_Ibcast}, etc.).

While the algorithm presented is pertinent both to parallel and distributed
computing, this work concentrates on MPI for parallel computing, where
the indicated collective operations have already been standardized and
are under wide use.  For MPI, checkpointing has at least three important
use cases.  Two first two use case are needed today:
(i)~chaining of long-running jobs across multiple allocations;
and (ii)~real-time or on-demand computing.  A THIRD, longer
range goal is system-level transparent checkpointing.

At HPC sites, the first MPI use case occurs for
long-running jobs needing multiple allocations
occur frequently in the case of applications
that exhibit only weak scaling.  (In strong scaling, doubling
the number of nodes assigned tends to almost halve the running
time.  This is not true for weak scaling.)
VASP~\cite{hafner2008ab} is a good example of a weak-scaling application.
VASP accounted for approximately 20\% of CPU time at the NERSC supercomputing center~\cite{NERSC} as of 2020~\cite[Figure~4]{driscoll2020automation}.
VASP supports multiple algorithms and data structures that are continually evolving.
As the VASP code evolves, it would be a large burden to continually update any application-specific module to reflect VASP's latest algorithms and data structures.

In many cases, library-based packages
may also be used to enable long-running
jobs with multiple allocations.
Examples of such packages include
VeloC~\cite{nicolae2019veloc}, SCR~\cite{moody2010design}, FTI~\cite{bautista2011fti},
ULFM~\cite{bland2013post,losada2020fault}, and Reinit~\cite{laguna2016evaluating}.
However, this assumes that the target application
satisfies a specific execution model:  a main loop, in which each iteration
is globally synchronized by an MPI barrier.
The model also assume that there are few or no active MPI objects
whose MPI context needs to be saved.  And finally, this
execution model assumes that there is not an overly long
wait for the main loop to return to the ``top'', where a
library-based checkpoint becomes possible.

The aforementioned VASP package is an example of a package with
a different execution model.  VASP is a multi-algorithm package
that can mix multiple numerical techniques.  It is for this reason
also that the VASP developers have not developed their own
application-specific checkpointing module~\cite{vasp2022martijn}, despite the obvious
need in cases of weak scaling.

The need for real-time and system-level checkpointing exposes another
motivation for transparent checkpointing.
A robust, transparent checkpointing package with a
common API, allows a sysadmin to validate compatible jobs in advance.
In essence, the sysadmin is ``given the keys'' to suspend ongoing jobs
without losing any already completed work.  Sysadmins are interested
for cases of emergency electrical shutdown,
unscheduled maintenance due to system intability, urgent external
computing requests (e.g., analysis of sudden astronomical events, output from large particle colliders),
and other urgent requests (see, for example,
XFEL~\cite{blaschke2021real,blaschke2023lightsource,giannakou2021experiences},
or examples from the workshop on \emph{Interactive and Urgent HPC} (\url{https://www.urgenthpc.com/}).

The history of transparent checkpointing is reviewed in Section~\ref{sec:relatedWork}.
In brief, BLCR~\cite{hargrove2006berkeley} (2006) had provided transparent checkpointing
on a single node.  This was briefly supported by MVAPICH~\cite{gao2006application}, Open~MPI~\cite{hursey2007design,hursey2009interconnect}, and DMTCP~\cite{cao2014transparent}
in order to support the original InfiniBand implementation.  But that support became inactive as HPC sites
moved on to newer network APIs, including HDR Mellanox InfiniBand,Cray GNI, Intel Omni-Path,
HPE Slingshot-10 and HPE Slingshot-11.

In the next phase, there was an effort to intervene at the MPI interface instead of the
network interface.  The MANA implementation of Barg \hbox{et al.} (2019)~\cite{garg2019mana} included a
two-phase-commit strategy for checkpointing, in which a barrier operation was inserted
before each collective call.  The barrier and following collective operation were
committed if all participating processes had reached the barrier at checkpoint time.
Otherwise, the barrier was aborted at checkpoint time.  However, as shown by a rich
set of experiments in Section~\ref{sec:experiments}, the overhead due to inserting
barriers at runtime was unacceptable.

This work presents a third checkpointing solution that uses multiple sequence numbers
to track and synchronize the ongoing collective calls without risking high runtime
overhead through additional runtime communication among participating processes.
Further, the original barrier-based algorithm of Garg \hbox{et al.} supported
only the blocking variant of collective calls.  Nonblocking collective calls
were not supported, as stated in~\cite[Section~4.2]{garg2019mana} (``Future Work'' section).

\subsection{Points of Novelty}

We introduce a new, low-overhead algorithm to efficiently support both blocking and non-blocking collective
communication.  In particular, non-blocking collective communication was one of the highlights of the new features introduced by MPI-3.0~\cite{message2015mpi}.  This work showcases three points of novelty.
\begin{enumerate}
\item The new algorithm greatly reduces the runtime overhead
      for collective operations, as compared to the older two-phase-commit
      algorithm.
\item This is the first transparent checkpointing algorithm that
      supports MPI's \emph{non-blocking} collective operations,
      needed for overlapping computation and communication.
\item The new algorithm is specially tested for scalability on the widely used
      VASP application. (VASP is responsible for approximately 20\% of CPU cycles
      at the NERSC supercomputing center).  VASP requires transparent
      checkpointing, since it does not include its own
      application-specific or library-based checkpointing package.  Thus, this work
      represents the first example of transparently checkpointing VASP,
      while maintaining low runtime overhead.
\end{enumerate}

\subsection{Organization of Paper}

The organization of this work is as follows.
Section~\ref{sec:background} provides brief background on MPI itself and the split-process architecture that forms the basis for MANA to checkpoint MPI.
Section~\ref{sec:mpi-standard} reviews some essential points in the semantics of MPI.
Section~\ref{sec:seq_num} presents a novel collective-clock (CC) algorithm, which replaces MANA's original two-phase-commit algorithm.
Section~\ref{sec:experiments} provides an experimental evaluation.
Section~\ref{sec:relatedWork} discusses the related work, and
Section~\ref{sec:conclusion} presents a conclusion.

\section{Background}
\label{sec:background}

Section~\ref{sec:mpi} reviews basic concepts for MPI itself, while Section~\ref{sec:2pcSplitProcess} reviews the \emph{split process} mechanism that is the basis for the MANA architecture.  The CC algorithm described here replaces the original two-phase-commit algorithm that was introduced in the original MANA paper~\cite{garg2019mana}.

\subsection{Review of MPI}
\label{sec:mpi}

Each MPI process has a unique rank as an MPI-specific process id.  MPI provides point-to-point operations on the ranks, such as \texttt{MPI\_Send} and \texttt{MPI\_Recv}. MPI also provides collective operations.  Examples include \texttt{MPI\_Barrier}, \texttt{MPI\_Bcast}, \texttt{MPI\_Alltoall}, etc.  A \emph{collective MPI operation} is executed in parallel by a subset of the MPI processes.  In this case, each member of the subset of processes for that operation must individually make a corresponding MPI call, to successfully invoke the parallel operation.

The subset of MPI processes participating in an operation is referred to as an \emph{MPI group}.  An \emph{MPI\_Communicator} can be created on top of an MPI group.  Creating a communicator is a parallel operation in which each participating process receives a handle to a common communicator representation, shared by all participating processes.
For a single MPI operation, all participating processes must call the same MPI collective function, and invoke the same MPI communicator.

An initial communicator \texttt{MPI\_COMM\_WORLD} includes all of the MPI processes.  Each new group numbers its processes consecutively, beginning with rank~0.
\texttt{MPI\_Group\_translate\_ranks} is available to determine the rank of a process within a new group, as compared to a previous group.

Finally, MPI also defines non-blocking variants, such as \texttt{MPI\_Isend}, \texttt{MPI\_Irecv}, \texttt{MPI\_Ibcast}, \texttt{MPI\_Ibarrier} and \texttt{MPI\_Ialltoall}.
 The variants have an additional argument, a pointer to an MPI \emph{request} object.
and does not block.

A non-blocking call first \emph{initiates} the MPI operation. MPI
immediately begins executing the operation. The non-blocking call immediately sets the request object and returns.
This enables overlap of communication and computation.
The MPI library \emph{globally completes} the  non-blocking operation soon after each participating process has locally initiated the corresponding call.

An individual MPI process tests if its part in the operation is \emph{locally complete} by calling \texttt{MPI\_Test}, \texttt{MPI\_Wait}, or a related call. 
The calls to test for completion include both the MPI request and a completion flag argument.
The current process has completed its part in the operation if the flag is set to true.
The request object is then modified in place to set its value to the pseudo-request \texttt{MPI\_REQUEST\_NULL}.

\subsection{MANA's Split Process Software Architecture and Checkpointing}
\label{sec:2pcSplitProcess}

The current work adopts MANA's \emph{split process architecture}.  That was introduced with
MANA (MPI-Agnostic Network-Agnostic transparent checkpointing)~\cite{garg2019mana}.
MANA is a plugin of DMTCP~\cite{ansel2009dmtcp}, which supports MPI applications using split processes.
Two processes are loaded into a single memory space. The upper half contains the
MPI application and a library of wrapper ``stub'' functions that redirects MPI calls to the lower half. The lower
half contains a proxy program that is linked with network and MPI libraries. On restart the
upper half is restored and the lower half is replaced by a new one. This design
decouples the MPI application from the underlying libraries that ``talk'' to the hardware.

When MANA takes a checkpoint, it saves only the memory regions associated with the
upper half.  When MANA restarts, it begins a new ``trivial'' MPI application with
the correct number of MPI processes.  Each MPI process becomes the new lower half,
and it restores to memory the upper-half checkpoint image file whose rank is the same
as the MPI rank in the lower half.

Since MANA does not save the lower-half memory, it works independently of the
particular network interface, and independently of the MPI implementation (providing
that the MPI implementation obeys MPI's standard API).

Thus, a ``safe'' state for MANA to checkpoint must obey the following invariant:
\begin{quote}
    \textbf{Collective Invariant:} \textit{No checkpoint may take place while an MPI process is inside a collective communication routine in the lower half.}
\end{quote}

The original MANA paper proposed a two-phase-commit (2PC) algorithm to find a safe state
obeying the collective invariant, above.
The core idea of the earlier two-phase-commit algorithm is to use a wrapper function
around each MPI collective call to insert a call to \texttt{MPI\_Barrier} (or a call
to \texttt{MPI\_Ibarrier} followed by a loop of calls to \texttt{MPI\_Test}).  When it is
time to checkpoint, if all processes have entered the barrier, then MANA waits
until all processes have completed the collective call.  If some processes have
not yet entered the barrier, then it is safe to checkpoint.  On restart, if one had entered
the barrier loop prior to checkpoint, then one again calls \texttt{MPI\_Ibarrier}
before continuing.

\section{A Close Look at the MPI Standard}
\label{sec:mpi-standard}

A prerequisite to understanding why the CC algorithm is correct requires
a precise reading of the MPI standard and its implications.
This document primarily cites the MPI-4.0
standard~\cite{message2021mpi}, the most recently published version.  But the same
quotes can be found in the MPI-3.0~\cite{message2015mpi} and MPI-3.1 versions.

There are two key points of information from the MPI standard, which will
be used frequently in this work.
\begin{quotation}
\noindent
First, a correct, portable MPI program must assume that MPI collective
operations (blocking and non-blocking) are synchronizing.
Any program violating this assumption is erroneous.
\end{quotation}

The reason is that for any given MPI collective operation, a particular
MPI library may implement the collective operation as synchronizing.
Hence, portable user programs must assume this more restrictive case.
In the words of the standard,
``a correct, portable program must invoke collective communications so that deadlock will not
occur, whether collective communications are synchronizing or not.''~\cite[Section~6.14]{message2021mpi}.

\begin{quotation}
\noindent
Second, once all participating MPI processes have initiated a non-blocking
operation, then the operation continues ``in background'', and must eventually
complete, independently of other actions by any of the MPI processes.
\end{quotation}

In the words of the standard,
``The progress of multiple outstanding non-blocking collective
operations is completely independent.''~\cite[Example~6.35]{message2021mpi}

\section{Collective Clock (CC) Algorithm}
\label{sec:seq_num}

The new algorithm is called CC due to its use of a happens-before relation on a vector of ``timestamps''.  In this sense, it bears a resemblance to the idea of logical clocks~\cite{baldoni2002fundamentals}.  The initial point of departure is that instead of employing logical clocks based on MPI processes, the \emph{collective clock} is based on logical clocks based on MPI communicators (in fact, on the underlying MPI groups).

The CC algorithm introduces a logical clock for each group of MPI processes, hereafter called a \emph{sequence number}.  The sequence number is initialized to zero.  When a collective operation occurs on a group, the sequence number for that group is incremented locally.  When a checkpoint (the analog of a distributed snapshot) is requested, a \emph{target (sequence) number} is computed for each MPI group, as the maximum of the sequence number of that MPI group that is seen at each MPI process.  For the purposes of target
numbers, two MPI groups are considered to be the same if they satisfy \texttt{MPI\_SIMILAR}, meaning
that they contain the same set of MPI processes.
(If an MPI process has never participated in that group for collective communication, then its sequence number is zero.)

An MPI process can then construct a local clock of target numbers.  A consistent snapshot can be obtained when each MPI process has reached the target number for each MPI process group with which it participates.

\subsection{Definitions}
\label{subsec:definitions}
Some formal terms are defined next, and used in the description of the algorithm.
\begin{description}
\item[\textbf{global group id (ggid):}] For the underlying MPI group of a communicator,
we assign a \emph{global group id (ggid)} based on hashing the ``world rank'' of each participating
MPI process according to its rank in MPI\_COMM\_WORLD. Communicator Ids generated by
the MPI library are only local resources handlers, so we need to compute the ggid to identify communicators
globally. By design, similar communicators in the sense of \texttt{MPI\_SIMILAR} have the same ggid.
\\
\emph{(NOTE: The computation of the world rank of a participating process is made
efficient by applying the MPI routine
\texttt{MPI\_Group\_translate\_ranks} (a local MPI operation)
to the local rank and MPI group for a given communicator.)}
\item[\textbf{Sequence number ({SEQ[ggid]}):}] The (local) sequence number of a ggid (often
denoted \emph{SEQ[ggid]})
is a local, per-process counter that
records the number of calls to (blocking) collective communication routines using
that MPI group.  If a particular MPI process has never invoked the given
MPI group, then the local sequence number for that ggid is zero.
\item[\textbf{Target number ({TARGET[ggid]}):}] The  (global) target number of a ggid
(often denoted \emph{TARGET[ggid]})
is a global value, representing the maximum of the local sequence numbers,
across each MPI process of the local sequence numbers, recorded within
that MPI process.
\item[\textbf{Reached a target}] A target is reached after we execute a blocking
collective call whose sequence number is equal to the target number.
\item[\textbf{Safe state}] A safe state of an MPI program in MANA is a state in execution
for which it's safe to checkpoint.
Two invariants must hold, to be in a safe state:

\textbf{Invariant 1:} \textit{No checkpoint must take place while a rank is inside a collective communication routine.}
This is the same as the collective invariant from Section~\ref{sec:2pcSplitProcess}.

\textbf{Invariant 2} \textit{If a collective communication call has started when the checkpoint request
arrives, then the checkpoint request must be deferred until
all members of the communication can complete the
communication.}
For example, suppose an \texttt{MPI\_Bcast} has started, and the sender process
has already broadcast its message.  Then the checkpoint
must be deferred for all receiving processes until they all can complete the communication.
\end{description}

\subsection{The CC Algorithm for Blocking Collective calls}
\label{sec:CCGeneral}

This subsection describes the CC algorithm in the context of blocking collective calls.  Later the interaction with blocking point-to-point calls will also be discussed.

Unlike MANA~\cite{garg2019mana}, the runtime overhead of the CC algorithm is almost zero
(see the micro-benchmarks of Section~\ref{sec:experiments}, since the only overhead is due
to interposing on MPI calls and incrementing
a sequence number.

\subsubsection{Overview of the CC algorithm}

When a communicator is created, if the ggid of the underlying group has
not yet been seen, then the sequence number of that ggid is initialized:
\texttt{SEQ[ggid]=0}.
During normal execution of an MPI application, each time a MANA wrapper
function is called on a blocking collective call, the \emph{ggid} for
that communicator is incremented.  Thus, if one participating process
has exited the blocking collective call and a second participating
process has either not yet reached that collective call or reached
it but not exited, then the value of \emph{SEQ[ggid]} in the
first process will be one more than the value in the second process.
These local operations are the only sources
of runtime overhead in the CC algorithm.

When  a checkpoint request arrives, execution continues under control of
the CC algorithm until a ``safe state'' is reached for all MPI processes.
First, all processes exchange all pairs (\emph{ggid, SEQ[ggid]}), where
\emph{ggid} is the global group id associated with a communicator known
to that process,
and where \emph{SEQ[ggid]} is the sequence number of that communicator.

Based on this exchange of information, each process sets \emph{TARGET[ggid]}
to be the maximum of the value of \emph{SEQ[ggid]} across all
processes where \emph{SEQ[ggid]} has been defined.
The CC algorithm then allows each MPI process to continue executing
until  \texttt{SEQ[ggid]==TARGET[ggid]} for each MPI process that
is a member of the global group id~\emph{ggid}.

\subsubsection{The complete CC algorithm (blocking calls)}

\begin{figure}[b!h!t]
\centering
\begin{subfigure}{0.41\linewidth}
\includegraphics[width=\textwidth]{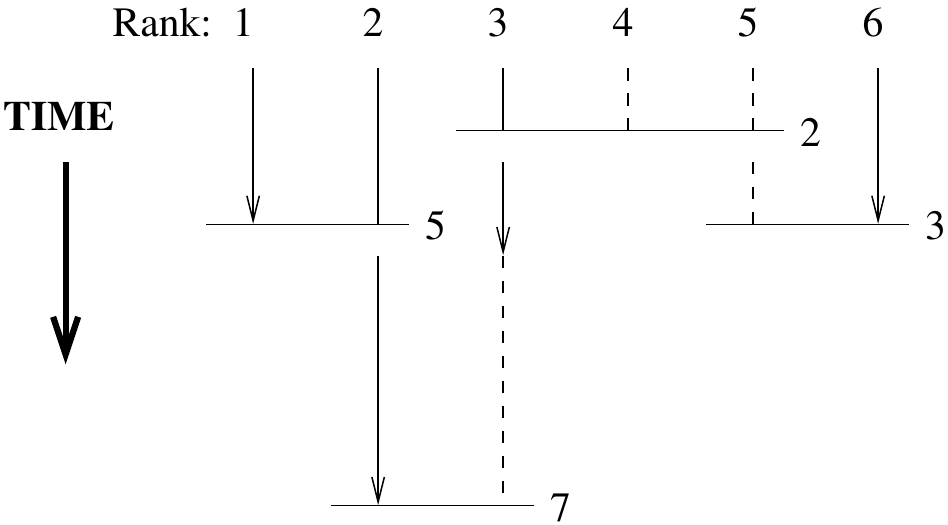}
\caption{Simple version of CC algorithm: setting the target sequence numbers
\\ \\ \\ \\ \\}
\label{fig:seq-num-algo-simple}
\end{subfigure}
\hspace*{0.1\linewidth}
\begin{subfigure}{0.41\linewidth}
\includegraphics[width=\textwidth]{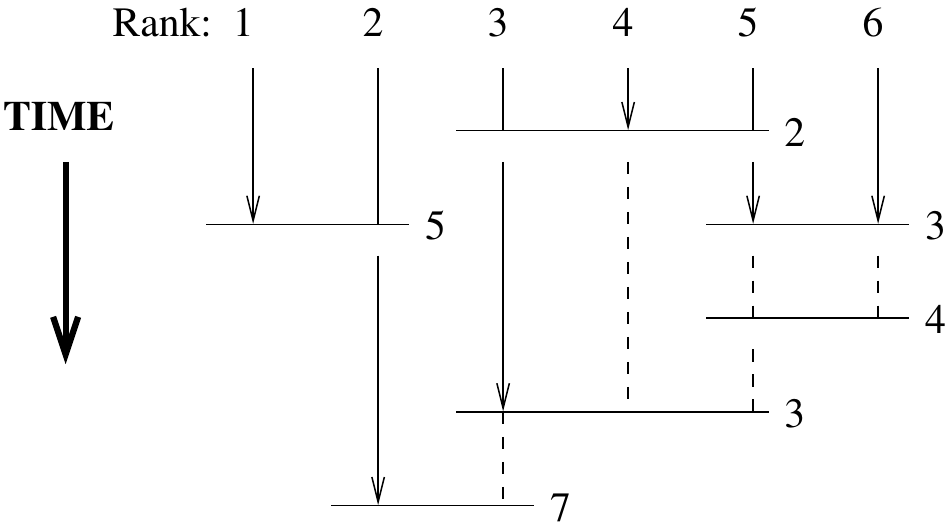}
\caption{Full version of CC algorithm (updating the target sequence numbers):
  When rank~3 encounters a new operation on group \{3,4,5\} (Case~2b), it increments
  the sequence number of the group to~3.}
\label{fig:seq-num-algo-full}
\end{subfigure}
\caption{\small The two figures are examples of
snapshots in time.
An arrowhead in the timeline of an MPI process in Figure~\ref{fig:seq-num-algo-simple}
indicates the current point in time at which the checkpoint request arrived.
A solid vertical line is in the past and a dashed vertical line is
in the future of the MPI process.  A dashed vertical line terminates
at the collective operation that is a target for the given rank.
A horizontal line indicates a (blocking) collective operation.
The number to the right of each collective operation is the sequence
number assigned for that ggid (for the set of ranks of the group of
that operation).}
\label{fig:seq-num-algo}
\end{figure}

Until now, we have assumed that in executing until reaching
a ``safe state'' for checkpointing, it is a simple matter to
share all \emph{SEQ[ggid]} across processes, compute
the \emph{TARGET[ggid]}, and wait until all targets
are reached.  However, the value of \emph{TARGET[ggid]}
is based on the values of \emph{SEQ[ggid]}
across all processes --- \emph{as known at the time of the
checkpoint request}.  Figure~\ref{fig:seq-num-algo} illustrates
an example causing \emph{TARGET[ggid]} to be
updated (increased), based on a collective operation that
is encountered \emph{only} after the checkpoint
is requested.

Figure~\ref{fig:seq-num-algo-simple} shows an example of target numbers.
The groups \{1,2\}, \{2,3\}, \{3,4,5\} and \{5,6\} (denoting ranks as
determined in \texttt{MPI\_COMM\_WORLD}) have
local targets 5, 7, 2, and 3, respectively.  Each individual rank may
have multiple targets.  For example, rank~3 has a target of~2 for
the group~\{3,4,5\}, and a target of~7 for the group~\{2,3\}.  Rank~3
has reached the target for the group~\{3,4,5\}, but not yet the
target for the group~\{2,3\}.

Figure~\ref{fig:seq-num-algo-full} shows a later snapshot.
All ranks have reached their targets, except for rank~3.
Rank~3 previously had a target of 7 for the group \{2,3\}.
But rank~3 encountered and just executed a new operation on group \{3,4,5\}.
Rank~3 increments its local sequence number, \texttt{SEQ[]}, for this group,
and determines that the local sequence number~3 is larger than
the previous shared target of~2 for that group.
(\texttt{SEQ[ggid])>TARGET[ggid]} for this group.)

Since the CC algorithm requires at all times that\linebreak[4]
\texttt{max(SEQ[ggid])==TARGET[ggid]} (for the maximum
over all MPI processes), the complete CC algorithm
adds a new step.  The newly incremented \texttt{SEQ[ggid]}
must be shared with all other participating processes.

So, rank~3
sends a message to ranks~4, and~5 with the new local sequence
number,~3, for the group.  Note that Rank~3 can locally
discover the peer ranks for the group \{3,4,5\} using the
local MPI call, \texttt{MPI\_Group\_translate\_ranks}.

Note also that since rank~5 has a new target for the group \{3,4,5\},
it will eventually execute the operation on group \{5,6\} again, and send
messages updating the target of that group to~4.  This will force
rank~5 to start executing again.

\subsubsection{The CC pseudo-code (blocking collective calls):}\hfill
\label{sec:pseudo-code}

It remains to show pseudo-code for the MPI function wrappers
used to implement the CC algorithm.

\begin{algorithm}[htb]
\caption{MANA's collective communication wrapper}\label{algo:wrapper}
\begin{algorithmic}
\Function{Wrapper}{$\ldots$, MPI\_Comm $comm$}
\State \Call{commit\_begin}{$comm$}
\State Call the real MPI function
\State \Call{commit\_finish}{$comm$}
\State \Return Return value of the real MPI function
\EndFunction
\end{algorithmic}
\end{algorithm}

Algorithm~\ref{algo:commit} shows the implementation
of \texttt{commit\_begin\(\)} and \texttt{commit\_finish\(\)}.

\begin{algorithm}[htb]
\caption{\texttt{commit\_begin} and \texttt{commit\_finish}}\label{algo:commit}
\begin{algorithmic}
\Function{commit\_begin}{MPI\_Comm $comm$}
\State \Call{wait\_for\_target\_updates}{ } // Algorithm~\ref{algo:wait}
\State $ggid \gets$ hash(MPI\_Comm\_group(comm, $\ldots$))\par
\State Increment sequence number $SEQ[ggid]$
\If{$ckpt\_pending$ {\bf and} $SEQ[ggid]$>$TARGET[ggid]$}
  \State $TARGET[ggid] \gets SEQ[ggid]$ 
  \State Send $TARGET[ggid]$ to other members of $ggid$ via MPI\_Isend
\EndIf
\EndFunction
\Function{commit\_finish}{MPI\_Comm $comm$}
\State \Call{wait\_for\_target\_updates}{ } // Algorithm~\ref{algo:wait}
\EndFunction
\end{algorithmic}
\label{algo:CC}
\end{algorithm}

Recall that the \emph{ggid} (global group id) is globally unique
across MPI processes, for the same MPI group.
The hash function computes this \emph{ggid}.
The MPI\_Isend/MPI\_Recv calls are used to update \emph{TARGET[ggid]},
as described earlier and shown in Algorithms~\ref{algo:commit} and~\ref{algo:wait}.
It is nonblocking to avoid deadlock.

The \texttt{wait\_for\_target\_updates} function is implemented by a loop of \texttt{MPI\_Iprobe} as shown in
Algorithm~\ref{algo:wait}. During a checkpoint, the \texttt{ckpt\_pending} flag will
be set to true. If all local target numbers are reached, the \texttt{commit\_begin\(\)} and
\texttt{commit\_finish\(\)} functions will cause the user thread to wait to see
if there will be further target updates.

The \texttt{MPI\_Iprobe} of Algorithm~\ref{algo:wait}
detects if there is any message from the MANA-internal communicator, mana\_comm, which is a duplicate of \texttt{MPI\_COMM\_WORLD}.
The communicator is used to share new target numbers for an application
communicator either if the communicator is new or if it is old and a new, larger target number has been found, during the process of reaching the global safe state.

\begin{algorithm}[htb]
\caption{wait\_for\_target\_updates}\label{algo:wait}
\begin{algorithmic}
\While{$ckpt\_pending$ {\bf and} all targets reached}
\State $flag \gets$ MPI\_Iprobe(MPI\_ANY\_SOURCE, mana\_tag, mana\_comm, ...)
\If{$flag = 1$}
\State $TARGET[ggid] \gets$ MPI\_Recv(..., 
\State \hbox{\ \ \ \ } status.MPI\_SOURCE, mana\_tag, mana\_comm, ...)
\EndIf
\EndWhile
\end{algorithmic}
\end{algorithm}

Until now, only the user (application) thread has been discussed.
The checkpoint thread also uploads the local sequence numbers seen
by each process to the key-value database of the DMTCP coordinator, in order
to determine a global value for the initial \emph{TARGET[ggid]} for each
communicator.

\subsubsection{A ``safe state'' is guaranteed}

In order to guarantee that we do not indefinitely insert new
communicators and increase the target number on some processes forever,
we introduce a ``happens-before'' relation and then
state a theorem, that there are no cycles in the happens-before relation.

A blocking collective operation~A \emph{happens before} operation~B if there is an MPI process participating in both operation~A and operation~B, for which operation~ is called before operation~B.  We extend the relation by transitivity:  If operation~A happens before operation~B and operation~B happens before operation~C, then operation~A happens before operation~C.  It also may happen that for two operations, neither operation completes before the other.

It remains to show that \emph{happens-before} is a well-defined relation.  We show that it cannot happen both that
operation~A \emph{happens before} operation~B and that operation~B \emph{happens before} operation~A.  Equivalently, this says that the \emph{happens-before} relation has no cycles.
\par\bigskip
\noindent
\textbf{Theorem:}  \emph{There is no cycle of blocking collective calls under the
  happens-before relation for blocking collective calls.}
\par\medskip\noindent
To demonstrate the theorem, we recall that blocking collective
calls can be assumed to be \emph{synchronizing} (see Section~\ref{sec:mpi-standard}).
Formally, ``an MPI collective procedure is \emph{synchronizing} if it
  will only return once all processes in the associated group or
  groups of MPI processes have called the appropriate matching MPI
  procedure.''~\cite[Section~2.4.2]{message2021mpi}

Next, given three blocking
collective calls, A, B and C, assume that A happens before B, and B
happens before C, and C happens before A.
Let A be called before B on process~X.
Let B be called before C on process~Y.
Let C be called before A on process~Z.
From the definition of syncronizing, these statement imply:
\begin{enumerate}
\item ``Y enters B'' only after ``X enters A''.
\item ``Z enters C'' only after ``Y enters B''.
\item ``X enters A'' only after ``Z enters C''.
\end{enumerate}
Clearly, this cycle is a contradiction.
The obvious generalization to a cycle of more than three blocking collective
calls then proves the theorem.\\
\emph{QED}

\bigskip
The theorem shows that there are no cycles.  In particular, note that
Algorithm~\ref{algo:commit} must update \emph{TARGET[ggid]} and send
update messages to the other participating processes of the \emph{ggid}
whenever \texttt{SEQ[ggid]\,>\,TARGET[ggid]}.  Since the collective
operation in question happens before an existing $TARGET[ggid']$
for the current MPI process, the number of new targets to be updated
due to the current process is bounded by the number of collective calls
before it reaches its targets.  New targets cannot be inserted
\hbox{ad infinitum}.

\subsubsection{Blocking collective calls and point-to-point calls}

\label{sec:CCBlockingP2P}

\begin{figure}[h!t]
\centering
\includegraphics[width=0.9\columnwidth]{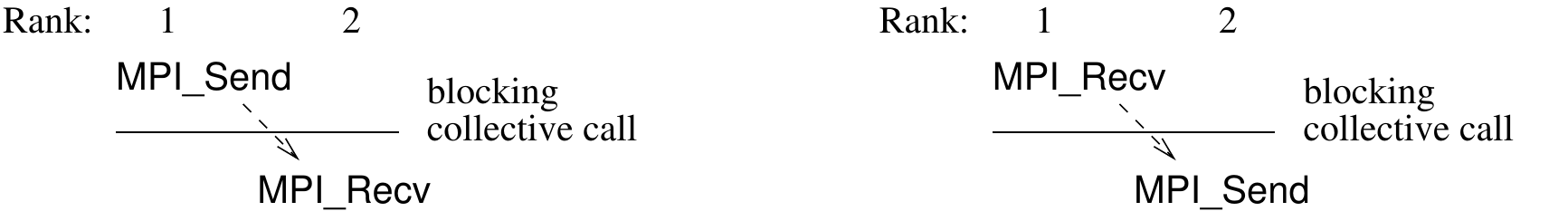}
\caption{The two cases above do \emph{not} occur in a correct MPI program.}
\label{fig:propertyB}
\end{figure}

 We recall again that blocking collective calls are assumed to be synchronizing (see
Section 3), and that ``an MPI collective procedure is synchronizing if it will only
return once all processes in the associated
group or groups of MPI processes have called the appropriate
matching MPI procedure.''~\cite[Section~2.4.2]{message2021mpi}.
Hence, a matched send-receive pair may not ``cross'' a blocking collective operation, as summarized in Figure~\ref{fig:propertyB}. 

\subsection{The CC algorithm extended to non-blocking calls}

It remains to describe the extension of the CC algorithm to non-blocking calls.
There are two issues to discuss:
\begin{enumerate}
    \item How are \emph{SEQ[ggid]} and \emph{TARGET[ggid]} set when a collective
    operation may be non-blocking?
    \item What should be done when a ``safe'' point is reached, but no processes have tested the operation
    for completion?
\end{enumerate}

\subsubsection{How are \emph{SEQ[ggid]} and \emph{TARGET[ggid]} set when a collective
    operation may be non-blocking}\hfill
\newline
The pseudo-code of Section~\ref{sec:pseudo-code} updates \emph{SEQ[ggid]} and \emph{TARGET[ggid]}
at the time of a blocking collective communication call.  However, in the non-blocking
case, this call is now split between a call that initiates a non-blocking collective communication
call (e.g., \texttt{MPI\_Ibcast}) and a call that completes the non-blocking collective
call (e.g., \texttt{MPI\_Wait} or \texttt{MPI\_Test}).

According to the MPI standard, non-blocking collective communication operations are independent from
other MPI operations (both blocking and non-blocking) after they are initiated.
Therefore, at any point of time between the initiation and completion of a non-blocking collective operation, 
the operation may or may not be executing in the background. The CC algorithm assumes all initiated nonblocking collective operations immediately
start executing in the background.  Therefore it increments the \emph{SEQ[ggid]} during the
initiation phase. This approach guarantees that all possible messages in the network are received before the safe state for checkpointing.

This choice to update during initiation, and not later, supports a common pattern. A process may initiate multiple non-blocking
collective communications at once, and waits for one or all processes to complete using functions like \texttt{MPI\_Waitany} and \texttt{MPI\_Waitall}.  (See~\cite[Example~6.35]{message2021mpi}.)

\subsubsection{What should be done if a ``safe'' point is reached,
but some processes haven't tested the operation for completion}
Since the CC algorithm increments the \emph{SEQ[ggid]} during the initiation of non-blocking communications,
it's possible that some of the communications haven't finish the communication when all processes reached all targets.

At a safe state, all processes of an incomplete non-blocking communication must have initiated the communication because of the invariant of a safe state.
Therefore, the communication will eventually complete if all processes start waiting for completion using functions like \texttt{MPI\_Test} and \texttt{MPI\_Wait}.
The CC algorithm keeps a list of \texttt{MPI\_Request} for incomplete non-blocking communications. When a safe state is reached, the CC algorithm will keep
calling \texttt{MPI\_Test} on each incomplete \texttt{MPI\_Request} until all communication has been completed.

\section{Experiments}
\label{sec:experiments}

All experiments were conducted on the Perlmutter Supercomputer at the National Energy Research Scientific
Computing Center (NERSC).  Perlmutter is the \#8 supercomputer on the Top-500 list
as of June, 2023~\cite{top500}.
Perlmutter has 3,072 CPU nodes and 1,792 GPU-accelerated nodes. Each CPU node has two AMD EPYC 7763 processors per node, for a total of 128~physical cores and 512~GB of RAM. The network uses Cray's Slingshot~11 interconnect.
Cray MPICH version is 8.1.25. Cray mpicc is based on gcc-11.2.
The Linux operating system is SUSE Linux Enterprise Server 15 SP4 (Release~15.4),
with Linux kernel~5.14. 

\subsection{Micro-benchmarks}

The OSU Micro-Benchmarks 7.0~\cite{network2022osu} were used to show the runtime overhead of the new CC algorithm compared to MANA's original two-phase-commit (2PC) algorithm.
Eight micro-benchmarks were chosen for different common patterns of blocking and non-blocking collective communications. Each experiment was repeated 5 times.

\begin{figure*}[h!t]
\centering
\begin{subfigure}{\textwidth}
   \centering
   \includegraphics[width=\textwidth]{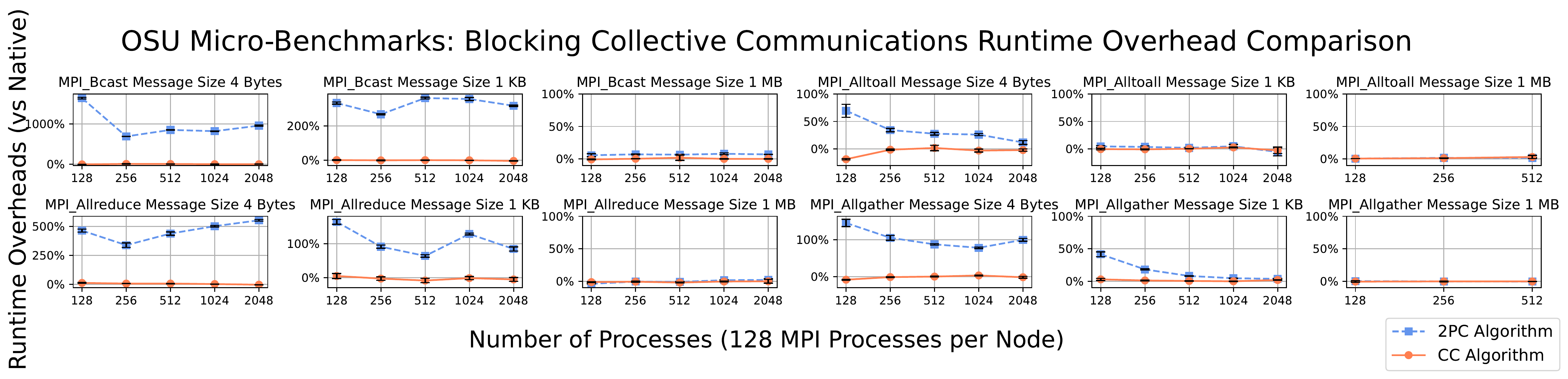}
\end{subfigure}
\begin{subfigure}{\textwidth}
   \centering
   \includegraphics[width=\textwidth]{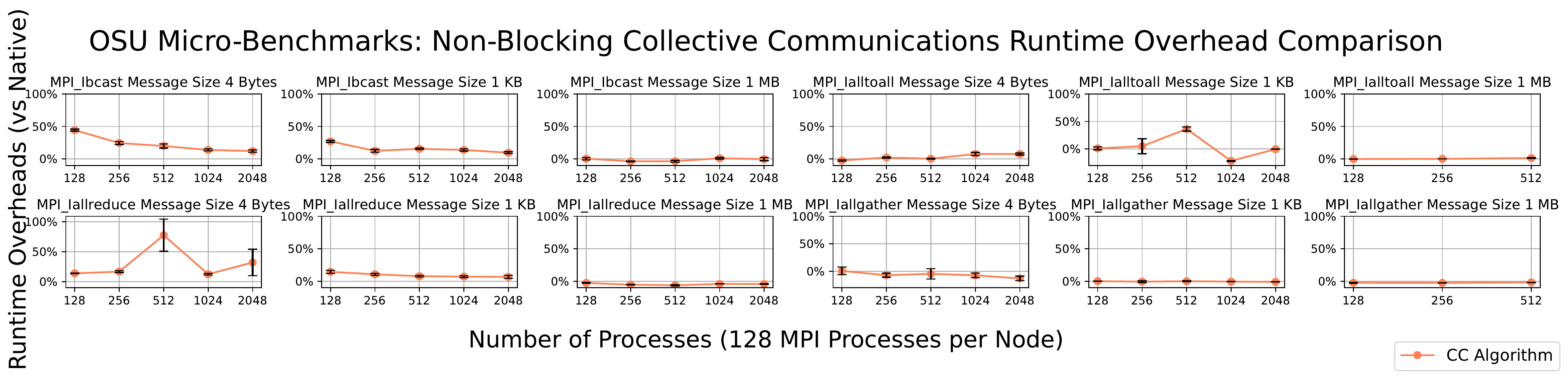}
\end{subfigure}
\caption{Runtime Overhead on Micro-Benchmarks for CC and 2PC.  Note that 2PC is not shown for non-blocking calls, since 2PC does not support such calls. In some cases, for example MPI\_Alltoall, both CC and 2PC algorithms have similar performance. So lines in the graph overlap with each other.}
\label{fig:micro_benchmark}
\end{figure*}

Figure~\ref{fig:micro_benchmark} shows the runtime overhead of both MANA's original two-phase-commit (2PC) and the CC algorithm of this work, with different message sizes (4~Bytes, 1~KB, and 1~MB). We scaled most micro-benchmarks up to 2048 processes over 16 nodes to test the scalability of the CC algorithm. The 2PC algorithm in the original MANA paper does not support non-blocking collective communication. Therefore, overhead is shown only for the CC algorithm, but not for 2PC.

MPI\_Alltoall/Ialltoall and MPI\_Allgather/Iallgather in the OSU Micro-Benchmarks do not support a message size of 1~MB over 1024 and 2048 processes, due to the default maximum memory limit. Hence, results are shown only up to 512 MPI processes for these cases.

\subsubsection{Blocking Collective Communication}
\label{sec:test_blocking}
In micro-benchmarks of blocking collective communications, the CC algorithm shows lower runtime overhead than 2PC, in most cases. In addition, the CC algorithm's runtime overhead remains consistently low as the message size or the number of nodes increases, whereas the runtime overhead of the 2PC algorithm varies depending on the message size and number of processes.

The 2PC algorithm inserts barriers that require extra communication and synchronization among processes. The additional communication increases the total latency of collective operations. Depending on the type of collective communication, the additional synchronization may have a different effect on performance. For example, MPI\_Bcast becomes slower because senders have to wait for all receivers to receive the message. But for functions like MPI\_Alltoall, the effect is minimal because the collective operation naturally requires synchronization among participating processors. 

In cases of large message size (1~MB), both algorithms perform identically to the native application. The cost of transferring messages is so large that the extra overhead introduced by each algorithm is insignificant.

\subsubsection{Non-blocking Collective Communication}
\label{sec:test_nonblocking}
The 2PC algorithm does not support non-blocking collective communications. Therefore, this section discusses runtime overheads only for the CC algorithm.

Note that for small messages, the runtime overhead for non-blocking communications is higher than for blocking counterparts. This is due to the communication being divided into two phases: initiation and completion. CC's wrappers for the two phases both contribute to runtime. So the constant runtime overhead becomes larger than for the single wrapper of a blocking call. Nevertheless, the runtime overhead quickly decreases as the message size and number of nodes increase.  

\begin{figure*}[h!t]
\centering
\centering
\includegraphics[width=\textwidth]{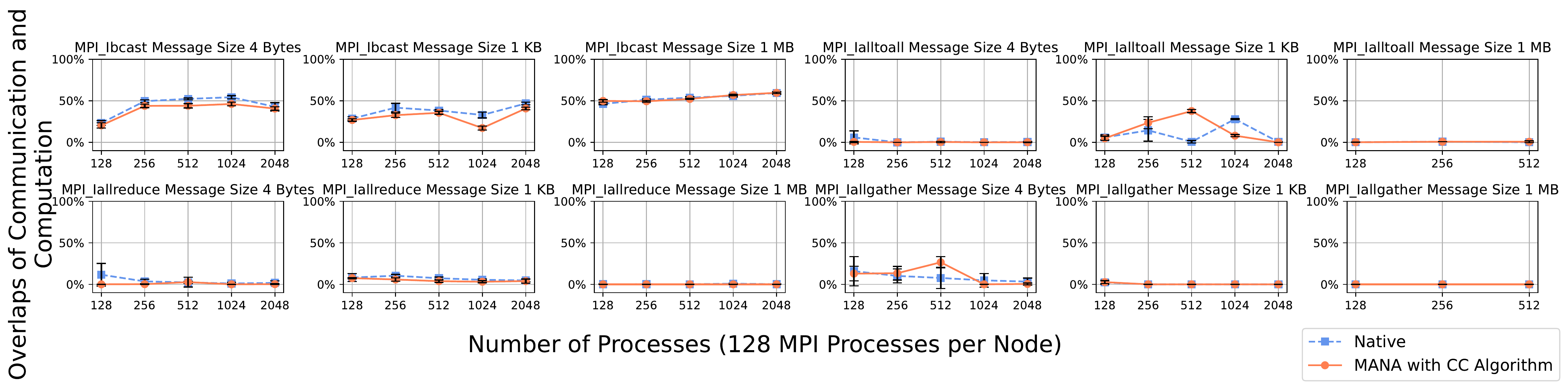}
\caption{Overlap of communications and computations when using non-blocking collective communications.}
\label{fig:overlaps}
\end{figure*}

Figure~\ref{fig:overlaps} shows the overlap of communication and computation for non-blocking collective communication. This overlap can improve the overall performance. The CC algorithm has a comparable amount of overlap, as compared to the native MPI implementation. Hence, the runtime overhead of CC in real-world programs is expected to be small, as seen in Sections~\ref{sec:real-world} and~\ref{sec:vasp}.

\subsection{Real-world Applications}
\label{sec:real-world}

\begin{figure}[h!t]
\centering
\includegraphics[width=0.8\columnwidth]{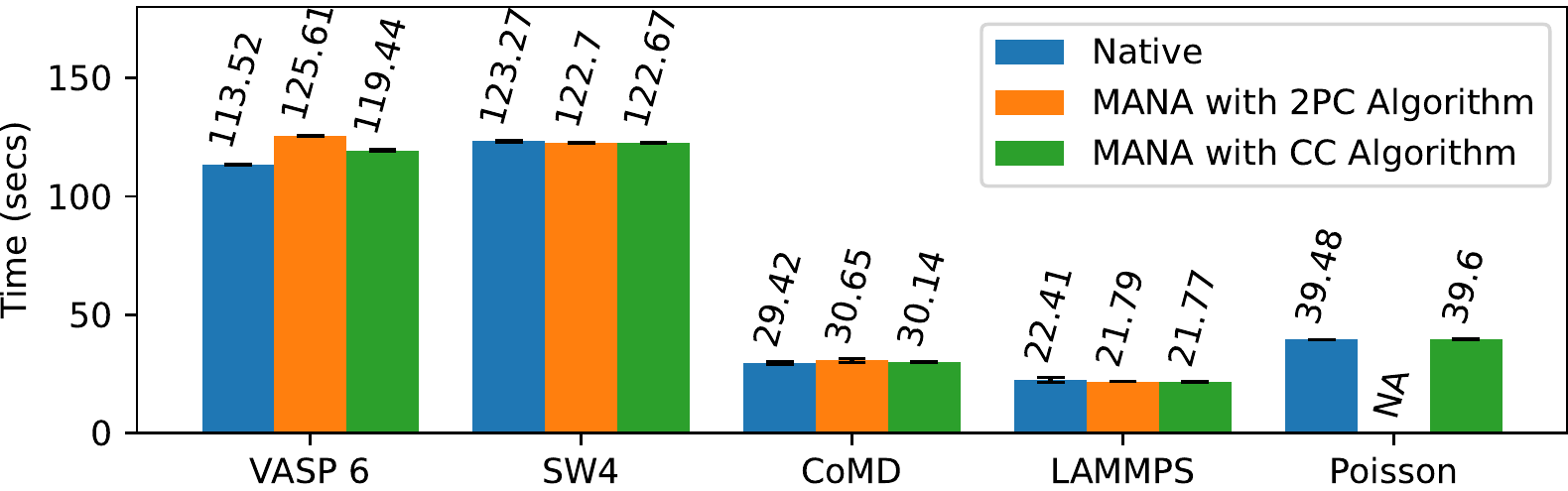}
\caption{Real-world applications runtime performance.}
\label{fig:real_world_overhead}
\end{figure}

Figure~\ref{fig:real_world_overhead} shows the runtime performance and standard deviation for five real-world applications.
Table~\ref{tbl:inputs} shows the experimental setup for each one, along with the rate of collective and point-to-point communication calls during the experiments. All experiments use 512 processes over 4~nodes on Perlmutter. Each test is repeated 5~times.

\begin{table*}[h!bt]
\centering
\begin{tabular}{ |l|c|c|r|r|}
\hline
\textbf{Application}  & \textbf{Processes} & \textbf{Input} &\textbf{coll\hbox{.} comm\hbox{.} calls} & \textbf{point-to-point  comm\hbox{.}} \\
                      &                    &                &\textbf{per second}                      & \textbf{calls per seconds} \\
\hline
VASP~6 & 512 & PdO4 & 2489.2 & 2568.9 \\ \hline
SW4 & 512 & LOH.1-h50.in & 0.6 & 157.9 \\ \hline
CoMD  & 512 & Cu\_u6.eam & 7.8 & 414.2 \\ \hline
LAMMPS & 512 & Scaled LJ Liquid & 6.3 & 1707.5 \\ \hline
Poisson Solver & 512 & rel error = 0.01 & 21.3 & NA \\ \hline
\end{tabular}
\caption{\label{tbl:inputs} Input for each application and communication rate.} 
\end{table*}

VASP~6 is one of the most commonly used applications at NERSC~\cite{driscoll2020automation}.It uses both collective communications and point-to-point communications intensively. Therefore, MANA shows the largest runtime overhead of all experiments. Nevertheless, the CC algorithm achieves a 5.2\% overhead under a heavy load. In comparison, the earlier 2PC algorithm has a runtime overhead of 10.6\%.

SW4~\cite{sjogreen2012sw4} uses little collective communication. Therefore, the CC and 2PC algorithms both have close-to-native performance. The average runtime shown in the graphs indicates MANA run slightly faster than the native application, which is within the standard deviation.

CoMD~\cite{papa2001constrained} and LAMMPS~\cite{thompson2022lammps} use more collective communications per second than SW4. However, the runtime overhead of MANA is still low.

The Poisson Solver~\cite{hoefler2007optimizing} uses non-blocking collective communications only. Therefore 2PC is not applicable. The CC algorithm achieves close-to-native performance.

\subsection{Scalability of Real-world Application: VASP}
\label{sec:vasp}

We show the CC algorithm's scalability beyond micro-benchmarks in the case of VASP~6. VASP~6 communicates intensively during the computation phase, which makes it a good example to test the scalability of our algorithm in an extreme, but still realistic scenario.  VASP~6 is tested with 128 processes up to 512 processes. 

\begin{figure}[h!t]
\centering
\includegraphics[width=0.6\columnwidth]{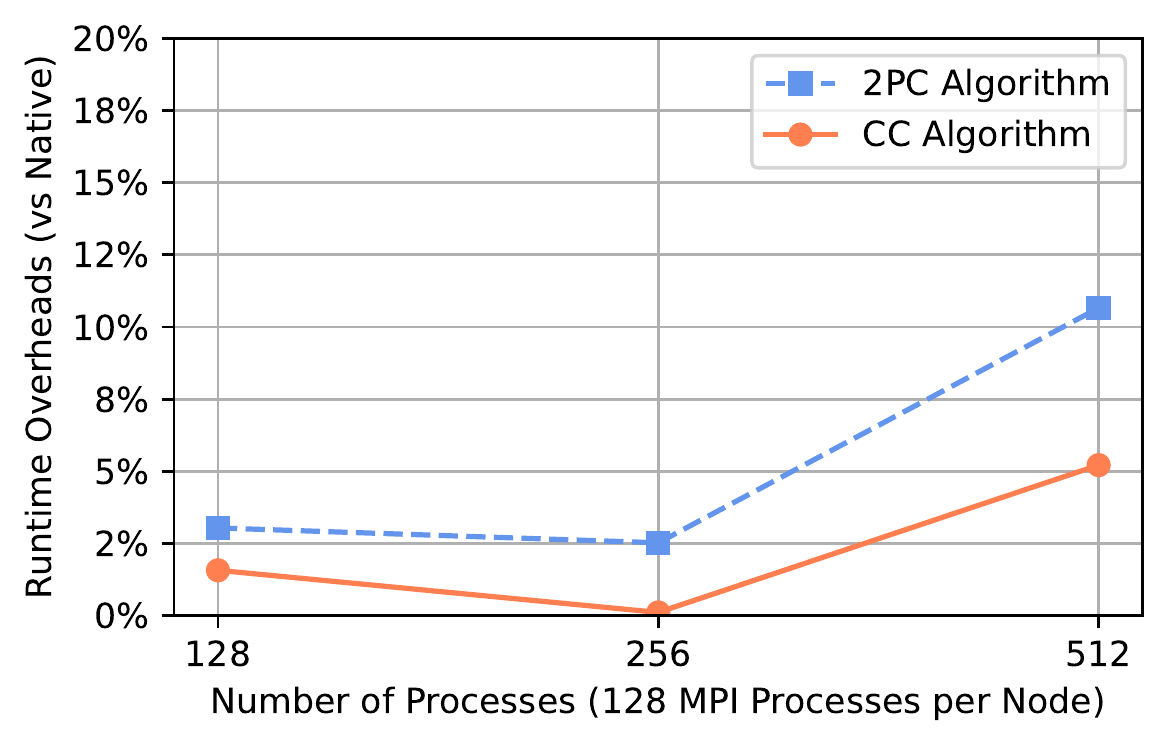}
\caption{VASP 6 Runtime Overhead Comparison}
\label{fig:vasp_overhead}
\end{figure}

Figure~\ref{fig:vasp_overhead} shows the runtime overhead in each case. The result shows that the CC algorithm scales better than the earlier 2PC algorithm in MANA. The remaining overhead is due to the point-to-point collective communications.  Point-to-point communication in MANA is out of the scope of this paper.

\section{Related Work}
\label{sec:relatedWork}

A rich set of library-based packages for checkpointing MPI exists:
SCR~\cite{moody2010design} (2010), FTI~\cite{bautista2011fti} (2011), ULFM~\cite{bland2013post,losada2020fault} (2014), and Reinit~\cite{laguna2016evaluating} (2016, a simpler interface inspired by ULFM), and VeloC~\cite{nicolae2019veloc} (2019).

Transparent checkpointing for MPI also has a long history.
It was demonstrated in MPICH-V~\cite{bouteiller2006mpich}.  Soon after, BLCR~\cite{hargrove2006berkeley} was created to provide transparent checkpointing for a tree of processes on a single computer node.  BLCR was then leveraged to support transparent checkpointing of MPI.  The strategy was for an individual MPI implementation to: (a)~disconnect the network connection; (b)~use BLCR to transparently checkpoint each individual node; and (c)~finally to reconnect the network (or connect it for the first time, if restarting from checkpoint images).  This was done for InfiniBand by MVAPICH~\cite{gao2006application}, for Open~MPI~\cite{hursey2007design,hursey2009interconnect}, and for DMTCP~\cite{cao2014transparent}.  Unfortunately, the underlying BLCR software, itself, was last updated only in January, 2013~\cite{blcrDownloads}, and with the newer networks, transparent checkpointing is no longer supported.

MANA~\cite{garg2019mana} was then developed in 2019, using both split processes and the two-phase-commit algorithm for collective calls.  The current work shows how to replace the two-phase-commit algorithm with more efficient collective clocks.

The CC algorithm can be viewed as a \emph{consistent snapshot algorithm} for MPI collective operations.  Consistent snapshots to support point-to-point operations in the case of distributed algorithms include:  the original Chandy-Lamport algorithm~\cite{chandy1985distributed} and  Baldoni \hbox{et al.}~\cite{baldoni2001rollback}.

Some example studies of fault tolerance for collective communication in distributed systems include Hoplite~\cite{zhuang2021hoplite}, scalable distributed collectives for Asynchronous Many-Task (AMT) models~\cite{whitlock2018scalable}, and atomic broadcast for Byzantine agreement~\cite{cristian1995atomic}.

\section{Conclusion}
\label{sec:conclusion}
The new CC algorithm reduces the runtime overhead for VASP from up to 10\% in MANA's old two-phase-commit (2PC) algorithm to typically 5\% or less in the new algorithm.  Similarly, the OSU Micro-Benchmark for blocking collective calls show a drastic improvement in this stress test (with up to 2048 processes):  from above 100\% runtime overhead with the old 2PC algorithm to nearly 0\% with the new collective-clock (CC) algorithm.  For non-blocking collective calls, the older 2PC algorithm does not support that case.  But the CC algorithm executes at a runtime overhead typically between 0\% and 10\%, for up to 2048 processes.
An atypical worst case occurs for \texttt{MPI\_Ibcast}, where CC can show runtime overheads of up to 50\%.

\bibliographystyle{alpha}
\bibliography{ref}

\newcommand{\etalchar}[1]{$^{#1}$}
\begin{thebibliography}{MBMDS10}

\bibitem[AAC09]{ansel2009dmtcp}
Jason Ansel, Kapil Arya, and Gene Cooperman.
\newblock {DMTCP}: Transparent checkpointing for cluster computations and the
  desktop.
\newblock In {\em 23rd IEEE International Parallel and Distributed Processing
  Symposium (IPDPS'09)}, pages 1--12. IEEE, 2009.

\bibitem[BBH{\etalchar{+}}13]{bland2013post}
Wesley Bland, Aurelien Bouteiller, Thomas Herault, George Bosilca, and Jack
  Dongarra.
\newblock Post-failure recovery of mpi communication capability: Design and
  rationale.
\newblock {\em The International Journal of High Performance Computing
  Applications}, 27(3):244--254, 2013.

\bibitem[BBP{\etalchar{+}}21]{blaschke2021real}
Johannes~P Blaschke, Aaron~S Brewster, Daniel~W Paley, Derek Mendez, Asmit
  Bhowmick, Nicholas~K Sauter, Wilko Kr{\"o}ger, Murali Shankar, Bjoern Enders,
  and Deborah Bard.
\newblock Real-time {XFEL} data analysis at {SLAC} and {NERSC}: a trial run of
  nascent exascale experimental data analysis.
\newblock Technical report, 2021.

\bibitem[BGTK{\etalchar{+}}11]{bautista2011fti}
Leonardo Bautista-Gomez, Seiji Tsuboi, Dimitri Komatitsch, Franck Cappello,
  Naoya Maruyama, and Satoshi Matsuoka.
\newblock {FTI}: High performance fault tolerance interface for hybrid systems.
\newblock In {\em Proceedings of 2011 international conference for high
  performance computing, networking, storage and analysis}, pages 1--32, 2011.

\bibitem[BHK{\etalchar{+}}06]{bouteiller2006mpich}
Aurelien Bouteiller, Thomas Herault, G{\'e}raud Krawezik, Pierre Lemarinier,
  and Franck Cappello.
\newblock {MPICH-V} project: A multiprotocol automatic fault-tolerant {MPI}.
\newblock {\em The International Journal of High Performance Computing
  Applications}, 20(3):319--333, 2006.

\bibitem[BHR01]{baldoni2001rollback}
Roberto Baldoni, Jean-Michel H{\'e}lary, and Michel Raynal.
\newblock Rollback-dependency trackability: A minimal characterization and its
  protocol.
\newblock {\em Information and Computation}, 165(2):144--173, 2001.

\bibitem[{BLC}]{blcrDownloads}
{BLCR team}.
\newblock Berkeley {L}ab {C}heckpoint/{R}estart for {L}inux ({BLCR}) downloads.
\newblock
  \url{https://crd.lbl.gov/divisions/amcr/computer-science-amcr/class/research/past-projects/BLCR/berkeley-lab-checkpoint-restart-for-linux-blcr-downloads/}.
\newblock "[Online; accessed Apr-2023]".

\bibitem[BR02]{baldoni2002fundamentals}
Roberto Baldoni and Michel Raynal.
\newblock Fundamentals of distributed computing: A practical tour of vector
  clock systems.
\newblock {\em IEEE Distributed Systems Online}, 3(02), 2002.

\bibitem[BWEB23]{blaschke2023lightsource}
Johannes~P Blaschke, Felix Wittwer, Bjoern Enders, and Debbie Bard.
\newblock How a lightsource uses a supercomputer for live interactive analysis
  of large data sets: Perspectives on the {NERSC-LCLS} superfacility.
\newblock {\em Synchrotron Radiation News}, pages 1--7, September 2023.

\bibitem[CASD95]{cristian1995atomic}
Flaviu Cristian, Houtan Aghili, Ray Strong, and Danny Dolev.
\newblock Atomic broadcast: From simple message diffusion to {B}yzantine
  agreement.
\newblock {\em Information and Computation}, 118(1):158--179, 1995.

\bibitem[CKAC14]{cao2014transparent}
Jiajun Cao, Gregory Kerr, Kapil Arya, and Gene Cooperman.
\newblock Transparent checkpoint-restart over infiniband.
\newblock In {\em Proc. of the 23rd Int. Symp. on High-Performance Parallel and
  Distributed Computing (HPDC'14)}, pages 13--24, 2014.

\bibitem[CL85]{chandy1985distributed}
K~Mani Chandy and Leslie Lamport.
\newblock Distributed snapshots: Determining global states of distributed
  systems.
\newblock {\em ACM Transactions on Computer Systems (TOCS)}, 3(1):63--75, 1985.

\bibitem[DZ20]{driscoll2020automation}
Benjamin Driscoll and Zhengji Zhao.
\newblock Automation of {NERSC} application usage report.
\newblock In {\em 2020 IEEE/ACM International Workshop on HPC User Support
  Tools (HUST) and Workshop on Programming and Performance Visualization Tools
  (ProTools)}, pages 10--18. IEEE, 2020.

\bibitem[GBBR21]{giannakou2021experiences}
Anna Giannakou, Johannes~P Blaschke, Deborah Bard, and Lavanya Ramakrishnan.
\newblock Experiences with cross-facility real-time light source data analysis
  workflows.
\newblock In {\em 2021 IEEE/ACM HPC for Urgent Decision Making (UrgentHPC)},
  pages 45--53. IEEE, 2021.

\bibitem[GPC19]{garg2019mana}
Rohan Garg, Gregory Price, and Gene Cooperman.
\newblock {MANA} for {MPI}: {MPI}-agnostic network-agnostic transparent
  checkpointing.
\newblock In {\em Proceedings of the 28th International Symposium on
  High-Performance Parallel and Distributed Computing}, pages 49--60, 2019.

\bibitem[GYHP06]{gao2006application}
Qi~Gao, Weikuan Yu, Wei Huang, and Dhabaleswar~K. Panda.
\newblock Application-transparent checkpoint/restart for {MPI} programs over
  {I}nfini{B}and.
\newblock In {\em Int. Conf. on Parallel Processing (ICPP'06)}, pages 471--478,
  2006.

\bibitem[Haf08]{hafner2008ab}
J{\"u}rgen Hafner.
\newblock Ab-initio simulations of materials using {VASP}: Density-functional
  theory and beyond.
\newblock {\em Journal of computational chemistry}, 29(13):2044--2078, 2008.

\bibitem[HD06]{hargrove2006berkeley}
Paul~H Hargrove and Jason~C Duell.
\newblock Berkeley {L}ab {C}heckpoint/{R}estart ({BLCR}) for {L}inux clusters.
\newblock {\em Journal of Physics: Conference Series}, 46(1):067, 2006.

\bibitem[HGLR07]{hoefler2007optimizing}
Torsten Hoefler, Peter Gottschling, Andrew Lumsdaine, and Wolfgang Rehm.
\newblock Optimizing a conjugate gradient solver with non-blocking collective
  operations.
\newblock {\em Parallel Computing}, 33(9):624--633, 2007.

\bibitem[HML09]{hursey2009interconnect}
Joshua Hursey, Timothy~I Mattox, and Andrew Lumsdaine.
\newblock Interconnect agnostic checkpoint/restart in {O}pen {MPI}.
\newblock In {\em Proc. of the 18th ACM Int. Symp. on High Performance
  Distributed Computing}, pages 49--58, 2009.

\bibitem[HSML07]{hursey2007design}
Joshua Hursey, Jeffrey~M Squyres, Timothy~I Mattox, and Andrew Lumsdaine.
\newblock The design and implementation of checkpoint/restart process fault
  tolerance for {O}pen {MPI}.
\newblock In {\em 2007 IEEE International Parallel and Distributed Processing
  Symposium}, pages 1--8. IEEE, 2007.

\bibitem[LGM{\etalchar{+}}20]{losada2020fault}
Nuria Losada, Patricia Gonz{\'a}lez, Mar{\'\i}a~J Mart{\'\i}n, George Bosilca,
  Aur{\'e}lien Bouteiller, and Keita Teranishi.
\newblock Fault tolerance of {MPI} applications in exascale systems: The {ULFM}
  solution.
\newblock {\em Future Generation Computer Systems}, 106:467--481, 2020.

\bibitem[LRG{\etalchar{+}}16]{laguna2016evaluating}
Ignacio Laguna, David~F Richards, Todd Gamblin, Martin Schulz, Bronis~R
  de~Supinski, Kathryn Mohror, and Howard Pritchard.
\newblock Evaluating and extending {U}ser-{L}evel {F}ault {T}olerance in {MPI}
  applications.
\newblock {\em The International Journal of High Performance Computing
  Applications}, 30(3):305--319, 2016.

\bibitem[MBMDS10]{moody2010design}
Adam Moody, Greg Bronevetsky, Kathryn Mohror, and Bronis~R De~Supinski.
\newblock Design, modeling, and evaluation of a scalable multi-level
  checkpointing system.
\newblock In {\em SC'10: Proceedings of the 2010 ACM/IEEE International
  Conference for High Performance Computing, Networking, Storage and Analysis},
  pages 1--11. IEEE, 2010.

\bibitem[{Mes}15]{message2015mpi}
{Message Passing Interface Forum}.
\newblock {MPI}: A {M}essage {P}assing {I}nterface standard: Version 3.1.
\newblock \url{https://www.mpi-forum.org/docs/mpi-3.1/mpi31-report.pdf}, June
  2015.

\bibitem[{Mes}21]{message2021mpi}
{Message Passing Interface Forum}.
\newblock {MPI}: A {M}essage {P}assing {I}nterface standard: Version 4.0.
\newblock \url{https://www.mpi-forum.org/docs/mpi-4.0/mpi40-report.pdf}, June
  2021.

\bibitem[NER]{NERSC}
{NERSC}, the primary scientific computing facility for the {O}ffice of
  {S}cience in the {U.S.} {D}epartment of {E}nergy.
\newblock \url{https://nersc.gov/}.

\bibitem[{Net}22]{network2022osu}
{Network-Based Computing Laboratory}.
\newblock Osu micro-benchmarks.
\newblock \url{https://github.com/forresti/osu-micro-benchmarks/}, 2022.

\bibitem[NMG{\etalchar{+}}19]{nicolae2019veloc}
Bogdan Nicolae, Adam Moody, Elsa Gonsiorowski, Kathryn Mohror, and Franck
  Cappello.
\newblock {VeloC}: Towards high performance adaptive asynchronous checkpointing
  at large scale.
\newblock In {\em 2019 IEEE International Parallel and Distributed Processing
  Symposium (IPDPS)}, pages 911--920. IEEE, 2019.

\bibitem[PMB01]{papa2001constrained}
Massimo Papa, Toshiki Maruyama, and Aldo Bonasera.
\newblock Constrained molecular dynamics approach to fermionic systems.
\newblock {\em Physical Review C}, 64(2):024612, 2001.

\bibitem[SP]{sjogreen2012sw4}
Björn Sjögreen and N.~Anders Petersson.
\newblock A fourth order accurate finite difference scheme for the elastic wave
  equation in second order formulation.
\newblock 52(1):17--48.

\bibitem[TAB{\etalchar{+}}22]{thompson2022lammps}
Aidan~P Thompson, H~Metin Aktulga, Richard Berger, Dan~S Bolintineanu,
  W~Michael Brown, Paul~S Crozier, Pieter~J in't Veld, Axel Kohlmeyer, Stan~G
  Moore, Trung~Dac Nguyen, et~al.
\newblock {LAMMPS}-a flexible simulation tool for particle-based materials
  modeling at the atomic, meso, and continuum scales.
\newblock {\em Computer Physics Communications}, 271:108171, 2022.

\bibitem[TOP23]{top500}
{TOP500}: The list.
\newblock \url{https://top500.org/}, June 2023.
\newblock [Online; accessed Oct-2023].

\bibitem[{VAS}22]{vasp2022martijn}
{Martijn Marsman}.
\newblock VASP developer.
\newblock Personal communication, 2022.

\bibitem[WKT{\etalchar{+}}18]{whitlock2018scalable}
Matthew Whitlock, Hemanth Kolla, Sean Treichler, Philippe P{\'e}bay, and
  Janine~C Bennett.
\newblock Scalable collectives for distributed asynchronous many-task runtimes.
\newblock In {\em 2018 IEEE International Parallel and Distributed Processing
  Symposium Workshops (IPDPSW)}, pages 436--445. IEEE, 2018.

\bibitem[ZLZ{\etalchar{+}}21]{zhuang2021hoplite}
Siyuan Zhuang, Zhuohan Li, Danyang Zhuo, Stephanie Wang, Eric Liang, Robert
  Nishihara, Philipp Moritz, and Ion Stoica.
\newblock Hoplite: Efficient and fault-tolerant collective communication for
  task-based distributed systems.
\newblock In {\em Proceedings of the 2021 ACM SIGCOMM 2021 Conference}, pages
  641--656, 2021.

\end{thebibliography}

\end{document}